\documentclass[aps,prd,
 ]{revtex4}
\usepackage{graphicx}
\usepackage{amsmath}
\usepackage{amssymb}

\newcommand\bef{\begin{figure}}
\newcommand\eef[1]{\label{fg:#1}\end{figure}}
\newcommand\beq{\begin{equation}}
\newcommand\eeq[1]{\label{#1}\end{equation}}
\newcommand\beqa{\begin{eqnarray}}
\newcommand\eeqa[1]{\label{#1}\end{eqnarray}}
\newcommand\fgn[1]{Figure \ref{fg:#1}}
\newcommand\eqn[1]{eq.\ (\ref{#1})}

\newcommand\ie{{\sl i.e.\/}}

\newcommand\Bsol{10^{57}}
\newcommand\bsol{A_\odot}
\newcommand\C{{\cal C}}

\newcommand\F{{\mathfrak F}}
\newcommand\fermi{E^F}
\newcommand\insb{\mathrm{in}}

\newcommand\msol{M_\odot}
\newcommand\PS{{\mathfrak S}}
\newcommand\R{{\mathfrak R}}

\newcommand\vm{\bar v}
\newcommand\zc{{\rm Z\/}_c}

\begin{document}

\title{Monsters and neutron stars}
\author{Sourendu\ \surname{Gupta}}
\email{sgupta@theory.tifr.res.in}
\affiliation{International Center for Theoretical Sciences,
 Tata Institute of Fundamental Research,\\ Hessarghatta Hobli,
 Bengaluru 560089, India.}
\begin{abstract}
Using dimensional arguments and other simple estimates, it is pointed out
that a branch of nuclei with mass number $A>10^7$ could be metastable
if they have close to maximal isospin. This branch is continuous with
neutron stars. The fastest decay mode of these hypothetical metastable
nuclei (which we call monsters) is through neutron evaporation. We
present an estimate of the lifetime and show it could be more than 100 ps
for $A>10^{39}$. We speculate about their creation in binary neutron star
mergers, where we argue that their lifetimes could be extended further,
and their presence could be detectable in gravity wave experiments of
the near future.
\end{abstract}
\maketitle

Stable nuclei are found in a range of mass numbers $A\le208$. All are
fairly close to being isoscalar (\ie, with proton number $Z\simeq
A/2$) with a maximum deviation less than 30\%. Around these stable
nuclei, one also has metastable nuclei with a variety of lifetimes
\cite{Majorana:2023ecz}. One definition of a metastable nucleus
which has been used is that its lifetime is greater than 100 ns
\cite{Kondev:2021lzi}. This is particularly stringent. Typical
$\beta$-decay lifetimes of hypernuclei are in the range of 200 ps,
and such nuclei can be resolved in experiments \cite{Lu:2026qgq,
HADES:2025qum}. This looser constraint may also be used to define
metastability.  Alternative definitions would change the size
of the island of metastability. The physics that limits this
island, whichever definition is used, involve strong and weak
interactions \cite{Pfutzner:2011ju} as well as electromagnetism
\cite{Pomeranchuk:1945a}).

However, there is also a branch of stable nuclear matter which we call
neutron stars (NSs). These have very large masses, of the order of a
solar mass ($\msol$), and have baryon number of the order of the baryon
content of the sun ($\bsol\simeq\Bsol$). NSs depart almost maximally
from isoscalarity, with $Z\ll A$. The stability of such nuclei involves
gravity. Here we argue that metastable nuclei could exist with properties
similar to NSs, albeit with smaller $A$. They would lie far outside the
neutron drip line, and hence will decay by neutron emission when gravity
does not stabilize them. We will call them metastable monster nuclei,
or monsters, due to the enormous values of $A$ that they have.

In order to understand why monsters might exist, it is useful to first
review the various limits on ordinary metastable nuclei and how NSs
evade these bounds. In this paper we will use natural units $\hbar=c=1$,
the symbols $m_e$, $m_n$ and $m_p$ for the electron, neutron and proton
masses respectively, and take the electron charge to be $e$. The nuclei,
including monsters, that we consider may be treated as liquid droplets,
so that the radius of an object with baryon number $A$ may be approximated
well by the empirical formula $a(A)=r_0\sqrt[3]A$ with $r_0=1.1$ fm.

We start by considering the limit set by a QED instability pointed out in
\cite{Pomeranchuk:1945a}. There it was shown that in a point-like nucleus
the Coulomb field is strong enough to destabilize the QED vacuum when the
nuclear charge exceeds a critical value $\zc= 137$. Later computations
showed that with more realistic charge distributions for nuclei this
instability sets in for $\zc=169$--173 \cite{Pieper:1969a, Popov:1970a,
Popov:1971c}. This then seems to gives an absolute upper bound for the
nuclear charge.  Although NSs have $Z/A\ll1$, $A$ is so large
that $Z$ exceeds this bound by many orders of magnitude. Nevertheless,
NSs are stable. The bound is evaded because electrons penetrate
inside the nuclear matter, so the material can be considered to be
locally charge neutral.

In order to extend this construction to finite nuclei, we need to argue
that local charge neutrality may continue to prevent the QED instability
even when $A\ll\bsol$. We may model a nucleus of baryon number $A$
as a sphere of radius $a(A)$ and charge $Ze$. Its interaction with an
electron has been examined using a Dirac equation with the potential
\beq
  V(r) = V_0\left[V_\insb\left(\frac ra\right)\,\Theta(a-r) 
	  -\frac ar\,\Theta(r-a)\right], 
  \qquad{\rm with}\qquad V_0=\frac{Ze^2}a,
\eeq{potential}
where $\Theta$ is the Heaviside step function. The outer potential is
Coulombic and due to a monopole charge $Ze$. We have factored an overall
scale $V_0$ from the inner potential, $V_\insb$. With this normalization,
continuity of the potential at the surface implies $V_\insb(1)=1$. If
the material is modeled as being uniformly charged, then one could
use classical electrostatics to show that $V_\insb$ is harmonic. The
QED instability with such a potential was examined numerically in
\cite{Czarnecki:2022ska}. Here we do not make such a strong assumption.

The potential of \eqn{potential} has two length scales, namely the nuclear
liquid drop size $a(A)$, and the inverse of the Coulomb energy scale
$a_\C(Z,A) = 1/V_0 = a(A)/(Ze^2)$.  The kinetic term in the Hamiltonian
has another length scale, namely the electron's Compton wavelength
$\lambda_e = 1/m_e$. One dimensionless parameter that one can build from
this is
\beq
  \PS = \frac a{a_\C} = Ze^2.
\eeq{strength}
This controls the strength of the coupling, and is clearly an important
parameter for the system. Another important dimensionless parameter 
is the ratio of the nuclear size and electron Compton wavelength,
which measures the finite size effect of the nucleus---
\beq
  \F = \frac a{\lambda_e} = am_e = 2.85\times10^{-3}\,A^{1/3}.
\eeq{size}
The Bohr-Rutherford model of an atom emerges in the limit $\F\to0$.
With three length scales we can define three dimensionless ratios,
although only two are independent.
\beq
   \R = \frac a{a_0} 
      = (Ze^2)\,(m_er_0A^{1/3}) \simeq 2.85\times10^{-3}A^{1/3}Ze^2.
\eeq{schroc}
This also vanishes in the Rutherford-Bohr limit. Note that $\R=\PS\F$.

\bef[tb]
 \includegraphics[scale=0.75]{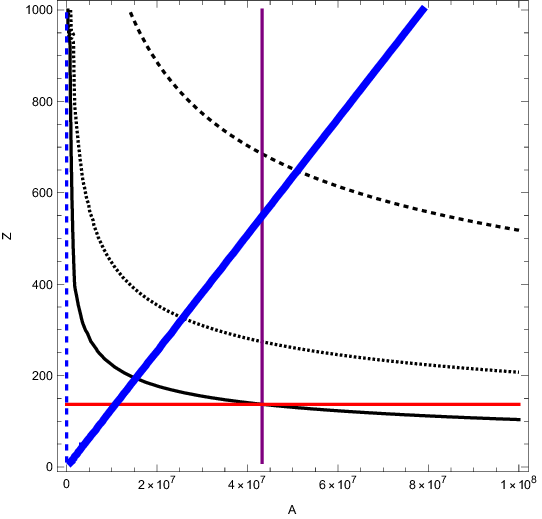}
\caption{In the plane of $A$ and $Z$ the vertical purple line corresponds
  to $\F=1$; to its right electrons may be localized within nuclear
  matter. The horizontal red line corresponds to $\PS=1$; below it there
  is no QED instability in the Coulomb problem. The lower left quadrant
  ($\F<1$ and $\PS<1$) is the domain within which normal nuclei are
  found; the upper right quadrant ($F>1$ and $S>1$) is the domain
  in which metastable monsters may exist. Also shown are contours of
  constant $\R$, for $\R=1$ (full line) $\R=2$ (dotted line), $\R=5$
  (dashed line). The thick blue line is the relation between $A$ and $Z$
  obtained for neutral, $\beta$-stable matter, see \eqn{protneut}. The
  dashed blue line is for isoscalar matter, and is almost vertical with
  this choice of scale.}
\eef{thompson}

Dimensional arguments can only give limited results, but some are
useful and interesting, especially when they are combined with our
detailed knowledge of the Bohr-Rutherford limit, $\F=0$. For example,
for the ground state energy of the electron, $E_e$, and the ground state
expectation of the square of the electron's radial position, $r_e$,
dimensional arguments allow us to write
\beq
  E_e=V_0\mathbb E(\PS,\F), 
  \quad{\rm where}\quad \lim_{\F\to0}\mathbb E(\PS,\F)=\frac\R2,
  \qquad{\rm or}\qquad
  V_0^2\left\langle r_e^2\right\rangle = \mathbb R(\PS,\F)
  \quad{\rm where}\quad \lim_{\F\to0}\mathbb E(\PS,\F)=\frac1{\F^2}.
\eeq{dimension}
For $\F=0$, textbook results for $E_e$ yield the Rydberg energy, $R =
V_0\R/2 = Z^2e^4m_e/2$.  Also, for $\F=0$ the RMS radius of an electron
is known to be the Bohr radius $a_0=1/(V_0\F)$.

A similar argument may be made for $\zc$, the lowest charge at which the
Pomeranchuk-Smorodinsky instability appears. Writing $\PS_c=\zc e^2$, one
sees that
\beq
  \PS_c=\mathbb S(\F), 
  \quad{\rm where}\quad \lim_{\F\to0}\mathbb S(\F)=1.
\eeq{vacbrk}
The $\F=0$ limit is the result of \cite{Pomeranchuk:1945a}. Further, the
numerical computations of \cite{Pieper:1969a, Popov:1970a, Popov:1971c}
for small $\F$ show that $\mathbb S$ increases with $\F$, \ie, the
curve $\mathbb S$ slopes up. From the limiting case of a NS, we know
that $\PS_c$ goes to infinity when $\F$ is large enough that the nuclear
charge is shielded because electrons are confined within the nucleus. This
can happen only when $\lambda_e<a$, \ie, $\F>1$, which implies that
$A>4.32\times 10^7$. This is, of course, only a necessary condition.
Dimensional analysis is unable to yield a sufficient condition on $\F$
for local electrical neutrality. So the value of $\F$ at which $\mathbb
P$ becomes asymptotically large requires a detailed study. What our
analysis shows is that there is a region of nuclear composition in which
Pomeranchuk-Smorodinsky instability could be avoided.

These points are summarized in \fgn{thompson} (notice the scales). QED
vacuum breakdown cannot occur in Bohr-Rutherford models for
$\PS<1$. Matter cannot be electrically neutral for $\F<1$. This quadrant
is the regime in which normal matter and normal atomic nuclei exist. QED
vacuum breakdown is avoided in the half plane with $\F>1$ due to matter
being locally electrically neutral. It should be noted that the region
of stability of normal nuclei is completely disjoint from the region in
which monsters may exist. We will next argue that monsters are also stable
against $\beta$-decay along the line marked with the thick blue line.

The stability of matter in a NS has been studied in detail including
nuclear and electroweak forces. For the moment we will assume that this
matter is put inside a confining potential. For a NS, gravity may be
thought of as providing this potential. We will return later to the
question of this potential for smaller monsters. Attractive (pairing)
forces between neutrons could then lead to condensates and Cooper-pairs
and superfluidity. However, for an examination of gross composition of
$\beta$-stable and locally neutral nuclear matter it seems possible to
model a cold NS as a Fermi liquid of its component neutrons, protons
and electrons. The Fermi energy, $\fermi$, is determined by the number
density, $n$, through the relation $n = (2m\fermi)^{3/2}/(3\pi^2)$,
where $m$ is the particle mass. Local charge neutrality requires the
proton and electron densities to be equal, which then gives the condition
\beq
 m_e\fermi_e = m_p\fermi_p.
\eeq{neutral}
The small ratio, $\epsilon=m_e/m_p$ implies that $\fermi_p \ll \fermi_e$.

For nuclei with more neutrons than protons, \ie, $Z<A/2$, weak
interactions provide an important decay mode. Since the neutrons
are filled to a higher energy than protons, the energy of the
nucleus may be lowered when a neutron transmutes into a proton through
a $\beta$-decay. Long-term stability of neutron stars demands that
$\beta$-decay and inverse $\beta$-decay be in equilibrium. Ensuring the
stability of a NS then involves matching the Fermi levels of all the
major components of the star, \ie,
\beq
 \fermi_n=\fermi_p+\fermi_e.
\eeq{beta}.
Solving this along with \eqn{neutral} gives $\fermi_p = \epsilon
\fermi_n/(1+\epsilon)$. The net densities of neutrons and protons are
\beq
  n_n = \frac1{3\pi^2}(2m_n\fermi_n)^{3/2}, 
      \qquad{\rm and}\qquad
  n_p = \frac1{3\pi^2}(2m_p\fermi_p)^{3/2},
      \qquad{\rm giving}\qquad
  \frac{n_p}{n_n} = \left(\frac\epsilon{1+\epsilon}\right)^{3/2},
\eeq{protneut}
using the vacuum relation $m_p\simeq m_n$.  Stability under QED
and weak decays together force nuclear matter to be almost maximally
non-isosinglet.  Nowhere in this treatment has gravity been used. So these
considerations apply to monsters as well as to NSs. The $Z/A$
ratio resulting from \eqn{protneut} has been shown in \fgn{thompson}.

Using the nuclear saturation density for $n_n$ in \eqn{protneut} along
with \eqn{beta} gives $\fermi_e \simeq \fermi_n = 63$ MeV, and $\fermi_p =
34$ KeV.  These numbers are surprisingly reasonable in spite of several
important approximations. The chief amongst them is the neglect of
interactions, whose main effect on this model would be to replace the
nucleon masses by effective in-medium values.  The computation also
indicates that electrons are mostly ultra-relativistic whereas neutrons
are at best mildly relativistic, so both components are more accurately
treated in relativistic kinematics than in the Newtonian approximation
used here. Protons, in contrast, are non-relativistic due to their
extremely small Fermi energy.

From \fgn{thompson} it is interesting to see that the
Pomeranchuk-Smorodinsky instability probably puts a lower bound to the
mass of a metastable monster. Since $\F\le1$ for $A\le 4.32\times 10^7$,
and the charge of a corresponding nuclide is likely to be above $\mathbf
S(1)$, it is quite likely that monsters smaller than this are forbidden.
So the branch of nuclei corresponding to metastable monsters is disjoint
from normal matter. We will see later that nuclear forces give stronger
bounds on the lower limit of metastable monsters.

The remaining bound on the islands of metastability of atomic nuclei are
set by the strong interaction. For a given $Z$, there are bounds on $A$
called the neutron drip line.  Locating the exact boundary of stability
remains an open experimental \cite{Baumann:2007jvt, Smith:2014noa}
as well as theoretical \cite{Ozawa:2000gj, Stoitsov:2003pd} problem.
Ongoing experiments are probing the limits of stability of nuclei,
and future facilities will test our theoretical understanding further.
This means that the confining potential which was assumed in \eqn{neutral}
and subsequently cannot be provided by internucleon forces. The whole of
the monster branch of matter lies well outside the neutron drip line.
If we try to build monsters with $A$ as small as $10^8$, then unlike
in a NS, gravity is negligible, and the monsters will decay due to
neutron drip.

Before examining this further, consider another route by which large
nuclei can decay, namely spontaneous fission.  This proceeds by
the deformation of an approximately spherical nucleus into two lobes
\cite{Bohr:1939ej, Strutinsky:1966bz, Nilsson:1969zz, Moller:1981zz}. For
normal nuclei the gain in surface energy may be compensated by a decrease
in the Coulomb energy as the two charged lobes separate out. However,
when matter is locally charge neutral, shape deformations only lead to
increase in surface energy without any compensating effect. Monsters
are therefore expected to be stable against fission.

Consider next the escape of single nucleons. The most common particle
being neutrons, one would expect that neutron evaporation could be the
dominant decay mode.  Since these are uncharged, one may set up a simple
model for the lifetime due to this process.  Assume that the monster
has initial mass number $A_0$ and, since neutrons evaporate from the
surface, the mass number after time $t$ is $A(t)$. Then one can write the
rate equation $\dot A = \Phi S$, where $\Phi$ is the flux of neutrons
coming to the surface and $S$ is the area of the monster. Treating a
monster as a liquid drop of neutral matter gives $S=4\pi r_0^2A^{2/3}$.
Deformations of this spherical surface could change the constant but
retain the $A^{2/3}$ dependence of $S$. In order to compute $\Phi$, we
note that the neutrons are densely packed inside the drop, and only those
within a distance $r_0$ of the surface may leave through the surface. If
they move with mean speed $\vm$ but at random angles, then only half
of them are correctly oriented to leave in time $r_0/\vm$. Denoting by
$d\dot n$ the rate at which neutrons leave an angular element $d\Omega$
of the surface in unit time, one can write
\beq
  d\dot n=n_n r_0\frac{d\Omega}{2r_0/\vm}, 
  \qquad{\rm implying}\qquad \Phi=\frac12n_n\vm.
\eeq{flux}
Assuming, as before, that this matter is at saturation density,
$n_n=3/(4\pi r_0^3)$, and that the Fermi velocity is $\vm=\sqrt{2
\fermi_n/m_n}$, we find
\beq
  \frac{dA}{dt} = \frac3{8\pi r_0^3}\,\sqrt{\frac{2\fermi_n}{m_n}}\,
     4\pi r_0^2A^{2/3}
    =\frac1\tau A^{2/3} \qquad{\rm with}\qquad 
    \tau=\frac{r_0}3\sqrt{\frac{2m_n}{\fermi_n}}.
\eeq{diffeq}
With the value of $\fermi_n=63$ MeV, quoted previously, this gives
$\tau\simeq2$ fm. This number could change if the condition for
$\beta$-stability is treated more accurately, but it would still be of
the order of a fm.

The solution to \eqn{diffeq} is
\beq
  A(t) = \left(\frac{t/\tau-\sqrt[3]{A_0}}3\right)^{1/3}
  \qquad{\rm giving\ the\ lifetime}\qquad
  t=\tau\sqrt[3]{A_0}.
\eeq{solve}
As required, $A(0)=A_0$.  Clearly this is an upper bound, since
faster processes may take over once $A$ becomes small enough. The
lifetime is plotted as a function of the mass of the monster in
\fgn{lifetimes}. Interestingly, the computation, which does not take
into account gravity, shows that a NS with baryon number $\bsol$ would
decay in a fraction of a second by neutron drip.

\bef[tb]
 \includegraphics[scale=0.75]{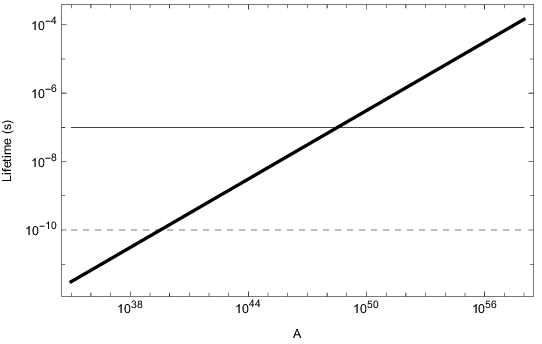}
\caption{Neutron drip lifetime of monsters, \eqn{solve} shown as a
 function of the starting mass number of the monster by the black diagonal
 line. The horizontal lines indicate two different definitions of the
 limits of metastability. The full line corresponds to a lifetime
 of 100 ns, and the dashed line to a lifetime of 100 ps. These two
 definitions give the limits of metastability to be $A>3.4\times10^{48} =
 3.4\times10^{-9}\bsol$ and $A>3.4\times10^{39} = 3.4\times10^{-18}\bsol$
 respectively.}
\eef{lifetimes}

For proton evaporation several factors suppress the decay rate by
decreasing the flux of protons at the surface. First, the fact that
$\fermi_p\ll\fermi_n$ decreases $v$, and since $n_p\ll n_n$ the
probability of finding protons at the surface again decreases. Taking
these two into account, we find that the proton evaporation rate is
controlled by the scale
\beq
 \tau_p = \tau\left(\frac{\fermi_n}{\fermi_p}\right)^2
    \simeq\frac\tau{\epsilon^2},
  \qquad{\rm giving}\qquad \tau_p\simeq6.7\times10^6\;{\rm fm\/}.
\eeq{protonevap}
Another factor which decreases the proton flux through the surface is
that an escaping proton would leave behind a charged monster, so costing
energy, which must be taken from the kinetic energy of the proton.
We could also take in the possibility that finite nuclear matter has a
neutron skin which is deficient in protons \cite{Fattoyev:2012rm}. This
would further suppress proton drip. All these indicate that proton
evaporation is a completely negligible process. Since proton densities
are so small, the probability of forming an $\alpha$ particle in a
monster are much smaller. As a result, decay of a monster through the
evaporation of light nuclei is negligible.

Let us consider in a little more detail the role of gravity in
stabilizing monsters.  The radius of a monster with baryon number
$\bsol$ in the liquid model is 11 Kms, with $r_0$ taken to be as
stated earlier. Furthermore, the radius is predicted to be 14 Kms
for a monster with $A=2\bsol$. These numbers are roughly consistent
with measurements and predictions for NSs of $\msol$ and $2\msol$.
While gravity certainly affects the radial density profile and details
of the structure, for an estimate of the radius of a NS at this level of
accuracy, the assumption of saturation density seems to be reasonable.
Assuming that a monster is an uniform ball of matter at saturation
density, one finds the escape velocity
\beq
 v_e(A) = \sqrt{\frac{2Gm_p}{r_0}}A^{1/3}.
\eeq{vescape}
This predicts $v_e(\bsol)=0.47$, which is consistent with estimates of the
escape velocity for a typical NS. A monster is stabilized by gravity when
$v_e$ is greater than the Fermi velocity used in \eqn{diffeq}.  Then using
\eqn{vescape} we find that monsters are stable for $A>4.6 \times10^{56}
= 0.46\bsol$. Lighter monsters are metastable up to the limits shown in
\fgn{lifetimes}. A more accurate estimate is worth investigating. This is
especially interesting now in view of recent papers which push the viable
lower bound of NSs to about $\msol$ \cite{Muller:2024aod, Salmi:2024bss}.

Metastable monsters cannot be constructed in a laboratory by bottom-up
fusion of smaller nuclei, since it would be impossible to move away
from approximate isoscalarity by this means. The huge gap in mass number
between the heaviest metastable normal nucleus and the lightest metastable
monster is another barrier to their synthesis in a lab. On the other
hand, there would be lower bounds to the mass of the NSs that can be
produced through supernova explosions. So an open question is how such
metastable monsters can be produced in the universe.

We propose that they might be produced in binary NS (BNS) mergers. BNS
mergers are events where the release of gravitational energy causes an
explosive kilonova. In this explosion a pulse of photons is released,
followed by a gas cloud of neutrons and light nuclei ejected from the
surface layers of the NSs. It is generally assumed that the remaining
material immediately merges into either a hot NS or black hole. However,
the existence of metastable monsters throws up the possibility that in
the final stage of BNS mergers a cloud of monsters can be formed. Lab
experiment show that the collision of two liquid drops produces a cloud
of small droplets as long as the kinetic energy of the drops is sufficient
to overcome the difference in surface energies. Since the energy produced
in such collisions is taken to be about 10\% of the mass of the system,
there is no energy barrier to the production of a cloud of metastable
monsters.  The analysis of the lifetime of isolated monsters presented
above may lead us to believe that even if these monsters are produced in
BNS mergers, they are irrelevant to astrophysics simply because of their
short lifetimes. We argue that next that there is reason to believe that
this is not true.

Consider that BNS mergers are likely to produce a very large number of
monsters. They, and neutrons evaporating from their surfaces, are bound
to each other through gravity since their total mass is the mass of the
merged system. This produces something like a fog of nuclear matter.
A fog is a two component state of matter in which droplets of liquid
(here the monsters) are immersed is a gas (which is made of evaporated
nucleons). Baryon number can be exchanged between the two components,
so one can write for the total baryon number $A_t=A_m+A_g$, where $A_m$
is the baryon number in the monsters, and $A_g$ in the gas.  If gas
can be lost from the fog then $A_t$ decreases with time.  

The number of monsters at a given time is $N$, and this number can either
remain constant or decrease, because droplets can evaporate away.  We will
also assume that the maximum distance between any pair of monsters is $R$,
so the fog lies inside a sphere of radius $R$ (see \fgn{circles}). We
will denote the surface area of this sphere by $S(R)$. Assuming that the
monsters are randomly distributed within this volume, the mean distance
between pairs is $d=R\sqrt[3]{4\pi/(3N)}$.

We denote the instantaneous baryon number of the $i$-th monster by $A_i$.
Then $A_m=\sum_i A_i$. The mean mass per monster is $A_m/N$, and we
denote the average surface area of a monster by $\overline S$.  If the
distribution of sizes of monsters is unimodal and does not have a long
tail, then one may expect that the approximation $\overline S=4\pi r_0^2
(A_m/N)^{2/3}$ does not have large corrections.

\bef[tb]
 \includegraphics[scale=0.7]{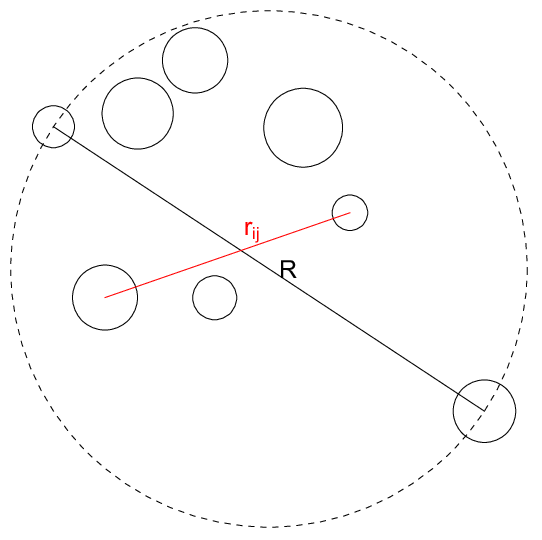}
 \caption{Illustrative figure showing some of the definitions used in
 computing the properties of a fog. The droplets of liquid are represented
 by the small circles. In this example the number of monsters is $N=8$.
 The distance between the $i$-th and the $j$-th monster is denoted
 $r_{ij}$. Its mean value is denoted by $d$ and its maximum by $2R$.
 The dashed outer circle denotes the containing sphere of radius $R$
 centered at bisector of the line joining the two monsters separated by
 distance $2R$.}
\eef{circles}

The outward flux at the surface of a monster is $\Phi$ as given in
\eqn{flux} and independent of $A_i$. At a distance $r$ it is attenuated to
$\phi_i(r)=\Phi S_i/r^2$. The net influx of nucleons at any point is
then
\beq
  \phi_{\rm in\/} = \frac{N\Phi\overline S}{8\pi d^2}
\eeq{influx}
Note that $S_i/(4\pi r_{ij}^2)$ is twice the angle subtended at the
position of the $j$-th monster by the surface of the $i$-th. This gives
a natural interpretation of \eqn{influx}. It is now easy to write the
evolution equations for $A_i$. These generalize \eqn{diffeq} by taking
care of inflow as well as outflow of nucleons
\beq
 \dot A_i = (\phi_{\rm in\/}-\Phi) S_i(n_i)
   =\left(\frac{N\overline S}{8\pi d^2}-1\right)\Phi S_i(n_i).
\eeq{coupled}
Note that the factor in brackets is the same for all monsters in a fog.
For an isolated monster, this factor is negative unity. Immersing a
monster in a fog clearly increases its lifetime.  Some of the droplets
which were unstable in isolation can become metastable.  In fact, if the
prefactor is positive, then the monster can even grow. This may happen,
for example, if there is an incoming flux of neutrons at the surface of
the bounding sphere of radius $R$.

Summing over all the monsters, one finds
\beq
 \dot A_m = \left(\frac{N\overline S}{8\pi d^2}-1\right)\Phi N\overline S.
\eeq{fogam}
There is a stable solution which corresponds to the prefactor vanishing.
Using the approximation $\overline S = S(A_m/N)$ along with this, one
may find a stable situation with
\beq
 A_m= \frac{8\sqrt2\pi}{3\sqrt N}\,\left(\frac R{r_0}\right)^{3/2},
\eeq{am}
where we used the relation given earlier between $d$ and $R$.
This situation can only arise if $\dot A_t=0$.

Since the remnant mass is large enough to gravitationaly bind the
fog, then one expects that in a time of order $\tau\sqrt[3]{A_m}$
one reaches the stable distribution of $A_m$ given in \eqn{am}. The
subsequent evolution can only be due to gravity.  The monsters with
masses stabilized by the fog continue to orbit each other, and re-merge
over time. This process may not be smooth, since the collision of two
monsters can again produce drops. However, the evolution of the fog will
produce gravitational waves. For a rough estimate of the frequency,
one may use the Keplerian time for a single monster in the CM of the
full system. This is not displaced much from the initial BNS signal,
and corresponds to the typical frequency range of current gravity wave
detectors. The important constraint is the amplitude of the signal.
From \eqn{am} one sees that $A_m\simeq\bsol/\sqrt N$. The chirp mass
of a single droplet in the center of mass of the fog therefore scales
as $N^{-3/10}$. The amplitude of the GW signal therefore scales as
$1/\sqrt N$. When $N$ is large the amplitude is too small to be detected.
However, successive re-mergers decrease $N$. As a result, in the late
stages of re-merger, when $N<100$, GW detectors of the near future may
be able to see the signals. The physics is interesting enough that these
gravitational signals are worth modelling.

{\bf Summary\/}: In this paper very simple estimates are given which
indicate the possibility of a second branch of metastable nuclei. They
are named monsters since they must have $A>10^7$ and isospin close
to maximal. Isolated monsters are stabilized only by gravitational
binding. From this consideration it is estimated that monsters with
$A>\bsol/2$ are stable, and therefore candidates for NSs. A plausible
estimate is that monsters with $A>10^{-18}\bsol$ may have lifetimes
greater than 100 ps. A consideration of the energy released in BNS mergers
shows that a fog of monsters might be created.  It is argued that in this
situation otherwise metastable monsters could become long lived. This
would give them time to re-merge into a remnant neutron star or black
hole. The end stages of this re-merger may be detectable by gravity wave
detectors of the near future. All arguments in this paper are qualitative,
rely on simple estimates, and hence are somewhat speculative. However,
the results seem to be interesting enough to pursue in detailed numerical
studies.

\end{document}